\documentclass[%
 reprint,
 amsmath,amssymb,
 aps,longbibliography
]{revtex4-2}
\usepackage{caption}
\usepackage{subcaption}
\usepackage{url}
\usepackage{graphicx}
\usepackage{dcolumn}
\usepackage{bm}

\begin{document}

\preprint{APS/123-QED}

\title{Absolute frequency measurements on the $5s5p^{3}$P$_{0}\to5s6d^{3}$D$_{1}$ transition in strontium}
\author{S. Zhang}
\email{zhangsn27@gmail.com}
\author{B. S. Tiwari}
\author{S. Ganesh}
\author{Y.Singh}
\email{y.singh.1@bham.ac.uk}
\affiliation{School of Physics and Astronomy, University of Birmingham, Edgbaston, Birmingham, B15 2TT, United Kingdom}
\author{V. V. Flambaum}
\affiliation{School of Physics, University of New South Wales, Sydney 2052, Australia}


\date{\today}

\begin{abstract}
We report the first absolute frequency measurements for the $5s5p^{3}$P$_{0}\to5s6d^{3}$D$_{1}$ transition at 394 nm for all the stable strontium isotopes by utilizing repumping induced spectroscopy in a magneto-optical trap. Absolute transition frequency is measured to be 760524409251(25) kHz for $^{88}$Sr. With reference to $^{88}$Sr, the isotope shifts are measured to be 91052(35), 54600(33) and 51641(28) kHz for $^{84}$Sr, $^{86}$Sr and $^{87}$Sr, respectively. We calculate the hyperfine constants A and B for the fermionic isotope $^{87}$Sr at kHz level. Furthermore, we perform King plot analysis by combining isotope shifts on the 689 nm transition to our data. 
\end{abstract}

\maketitle

\section{\label{sec:level1}Introduction} 
Owing to their favorable atomic properties, alkaline-earth-metal (-like) atoms, for example, Ca, Sr (Yb), have excellent suitability for various applications ranging from atomic optical clocks \cite{54,55,56,62,63,74,64,75}, optical tweezer-based quantum simulation and computing \cite{65,68,69,70,71}, to Sisyphus cooling technology \cite{66,67,73}, and to dipole-dipole interactions for many body physics\cite{57,72}. Moreover, with the recent advancement of quantum technologies these atoms are playing a more significant role than ever before \cite{90,91,92,93,94}. The atoms have emerged as excellent tools to probe new fundamental interactions using isotope shift spectroscopy \cite{98,99,100,101,102}. Among them, strontium with four stable isotopes, is uniquely placed for carrying out such a fundamental work. Strontium isotopes have either most spherical ($^{88}$Sr) or close to near spherical nuclei due to their proximity to magic number of neutrons (magic number 50). This ensures that nuclear deformation as a main source of non-linearity in a King plot is non-existent in Sr\cite{103,104}. Nuclear deformation is the main source of non-linearity in heavy atoms with deformed nuclei like Yb. Also, due to smaller charge of Sr, field shift which is sensitive to nuclear charge distribution, is much smaller in Sr than in heavier elements like Yb. In case of Sr compared to Ca, quadratic effects in mass shifts are also smaller. In such elements, transitions between singlet and triplet states feature ultranarrow lines and long coherence which are extremely beneficial in particular for optical clocks. So far, transitions between triplets and singlets have caught much attention, in contrast, transitions between triplets are less extensively researched, although the significance of these transitions cannot be ignored on numerous grounds.  

The transitions originating from $5s5p^{3}$P$_{0,1,2}$ to $5snd^{3}$D$_{1,2,3}$ can act as a channel to repump atoms in $^{3}$P$_{2}$ and $^{3}$P$_{0}$ back to the cooling cycle of the $^{1}$S${_0}\to^{1}$P${_1}$ magneto-optical trap (MOT) \cite{76,77,82}. The $^{3}$P$_{2}\to^{3}$D$_{3}$ midinfrared transition at 2.9 $\mu$m of Sr is an alternative for MOT which enables to relax the requirement of laser frequency stabilization in contrast to the 689 nm transition \cite{95,96}. Precise knowledge of the transitions $5s5p^{3}$P$_{0}\to^{3}$D$_{1,2,3}$ is necessary for the calculation of the $^{3}$P$_{0}$ polarizability\cite{81}. Experimental measurements of $^{3}$D$_{1}$ decay rate to $^{3}$P manifold can improve the largest systematic uncertainty - blackbody radiation (BBR) dynamic coefficient in optical lattice clocks\cite{78,79,80,83,97}. The $5s5p^{3}$P$_{0}\to5s4d^{3}$D$_{1}$ at 2.6 $\mu$m of bosonic strontium confined in a dense array is of great interest for the study of dipole-dipole interactions\cite{84,85,89}, topological quantum optics\cite{86,88} and quantum electrodynamics\cite{87}.
To determine the frequencies of these transitions, the method of precise spectroscopy is applied to a cold atomic sample. The frequencies of $5s5p^{3}$P$_{2}\to5snd^{3}$D$_{1,2,3}$ are measured using reservoir spectroscopy to 5 MHz accuracy for all the stable isotopes and 200 kHz uncertainty for isotope shifts \cite{13}. Another report on the frequency measurement of $5s5p^{3}$P$_{2}\to5s4d^{3}$D$_{2}$ mentions 80 MHz uncertainty by using the electron shelving technique \cite{19}. More recently, the frequency of $5s5p^{3}$P$_{2}\to5s4d^{3}$D$_{3}$ is measured again for $^{88}$Sr with an uncertainty of 9 kHz, improved by nearly three orders of magnitude using the photon-momentum-transfer technique \cite{6}. However, the transitions originating from $5s5p^{3}$P$_{0}$ to the higher excited states $5s6d^{3}$D$_{1,2,3}$ have not yet been investigated.

In this paper, we focus on the $5s5p^{3}$P$_{0}\to5s6d^{3}$D$_{1}$ transition due to the following two main reasons. Firstly, the transition is ideal for repumping of the $^{1}$S$_{0}\to^{1}$P$_{1}$ MOT; secondly, this transition can be used as an imaging channel for the 2.6 $\mu$m $5s5p^{3}$P$_{0}\to5s4d^{3}$D$_{1} $ transition which can be used to study long-range dipole dipole interactions \cite{57}. In our case, we employ repumping induced spectroscopy (RIS) \cite{13,19,33,18} to measure an absolute frequency of $5s5p^{3}$P$_{0}\to5s6d^{3}$D$_{1}$ for the first time for all the stable Sr isotopes. A 394 nm laser is shone to atoms cooled in a continuous steady-state 461 nm MOT with a 707 nm laser pumping atoms in $^{3}$P$_{2}$ back to the cycle. The atom number is enhanced while the 394 nm laser is frequency-scanned to the resonance and the emitted fluorescence is collected by a camera (Andor Zyla 5.5). The absolute frequency is measured by linking the 394 nm laser to a 698 nm clock laser via a transfer cavity. The clock laser frequency is read out by a frequency comb (MenloSystems SmartComb). By taking the lead frequency shift contribution, the measurement uncertainty is obtained to be less than 30 kHz for all the stable isotopes. The isotope shifts can then be derived by referencing other isotopes to $^{88}$Sr. For the measurement of $^{87}$Sr, the two hyperfine constants A and B are determined. In addition, we perform King plot analysis by combining our data and the 689 nm transition data from the reference \cite{27}.

\begin{figure*}
\centering
\includegraphics[width=14 cm]{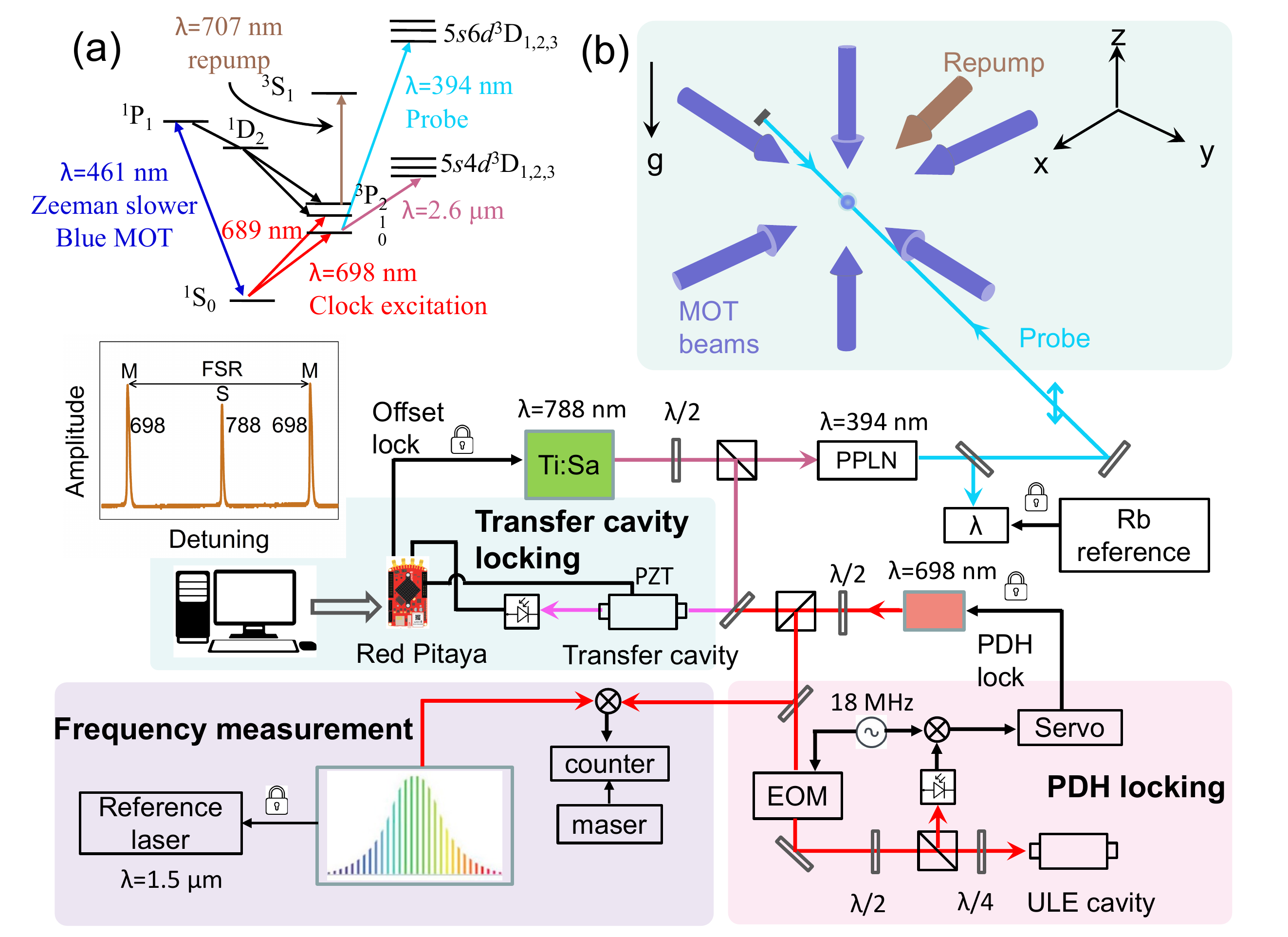}
\captionsetup{justification=raggedright}
\caption{\label{fig:setup} Schematic diagram of the experiment. (a) Relevant energy levels of Sr atom. We perform precision measurements on the $5s5p^{3}$P$_{0}\to5s6d^{3}$D$_{1}$ transition at 394 nm. (b) Schematic illustration of experimental setup. The whole setup mainly consists of four parts, i.e., transfer cavity locking, PDH laser stabilization, cold atomic package and frequency comb measurements. The probe laser is generated by frequency doubling 788 nm laser locked to an ultrastable clock laser at 698 nm via a transfer cavity with the help of a programmable micro controller (Red Pitaya). The clock laser is stabilized to a vertical cavity with a finesse of 2.3$\times$10$^5$, and has a 100 Hz linewidth and 1.1$\times$10$^{-14}$ stability at 1 s. Ti:Sa: Ti:Sapphire; EOM: electro-optic modulator; PPLN: Periodically Poled Lithium Niobate. Our frequency comb is referenced to an ultrastable 1.5 $\mu$m laser with stability of 2.2$\times$10$^{-15}$ at 1 s. All the frequency measurements are referenced to a hydrogen maser. The atomic ensemble is trapped in a MOT at $^{1}$S$_{0}\to^{1}$P$_{1}$ transition.}
\end{figure*}
\section{Experimental Details}
A schematic of the experimental setup and relevant energy levels for $^{88}$Sr are shown in Fig.~\ref{fig:setup}. Sr atoms with a flux of $10^{11}$s$^{-1}$ ejected from an oven are decelerated by a Zeeman slower and captured in a three-dimensional (3D) magneto-optical-trap (MOT) operated on the broad line $^{1}$S$_{0}\to^{1}$P$_{1}$ at 461 nm. The MOT beams, red-detuned by 40 MHz, $\texttt{i.e.}$, 1.25$\Gamma$ ($\Gamma$=32 MHz), have a total peak intensity of 0.5I$_{s}$ and a $e^{-2}$ radius of 1 cm under the operational condition. The axial magnetic gradient is 55 G/cm. We pump atoms populated in $^{3}$P$_{2}$ by 707 nm light back to the cooling cycle through the $^{3}$P$_{2}\to^{3}$S$_{1}\to^{3}$P$_{1}\to^{1}$S$_{0}$ channel. The repump laser is locked to a high precision wavemeter (highfinesse WSU2) with an accuracy of 2 MHz. Driving the $5s5p^{3}$P$_{0}$-$5s6d^{3}$D$_{1}$ transition at 394 nm enables to enhance the atom number in the steady-state MOT by an order of magnitude in comparison to the repumping-free case, which is the idea of RIS. 

We carry out RIS to measure the frequency of the $5s5p^{3}$P$_{0}\to5s6d^{3}$D$_{1}$ transition for all the stable isotopes. The procedure of the measurement is briefly summarized as follows. Sr atoms are continuously loaded into a $^{1}$S$_{0}\to^{1}$P$_{1}$ MOT following Zeeman slowing and with the 707 nm repumper for the $^{3}$P$_{2}$ atoms. As such, the atom number in the MOT stays stable under the working conditions unless another repumper at 394 nm brings the $^{3}$P$_{0}$ atom back to the cooling cycle via the $^{3}$P$_{0}\to^{3}$D$_{1}\to^{3}$P$_{1}\to^{1}$S$_{0}$ channel. The 394 nm probe laser then illuminates the MOT atoms at a low power, typically at 70 $\mu$W, leading to an increased atom number in the MOT. The atomic fluorescence is collected and detected by the camera. The counterpropagating configuration is employed for the probe laser to cancel the first-order Doppler shift. For measuring the transition frequencies of various isotopes, the MOT is tuned to work at the respective frequencies of $^{1}$S$_{0}\to^{1}$P$_{1}$. At resonance, the MOT atom number is enhanced by a  factor of four with the 394 nm repumper at 70 $\mu$W on.

Pumped by a high-power fibre laser at 532 nm, the probe laser has a frequency tunability of hundreds of GHz. As we have no lab-ready frequency comb to beat with, the frequency of the probe laser is determined by the seed laser frequency at 788 nm referenced to an ultrastable clock laser at 698 nm through the transfer cavity locking scheme \cite{17,31}. The clock laser is stabilized to an ultralow-expansion glass (ULE) cavity with a finesse of 2.3$\times$10$^{5}$ by means of Pound-Drever-Hall (PDH) technique, resulting in a stability of 1.1$\times$10$^{-14}$ at 1 s. The frequency of the clock laser can be directly measured by a frequency comb (Menlo Systems Smart comb) which has a repetition frequency $f_{\texttt{rr}}$=125 MHz and a carrier envelope offset frequency $f_{\texttt{ceo}}$=10 MHz, referenced to an ultrastable laser at 1542 nm with a stability of 2.2$\times10^{-15}$ at 1 s. The beat note between the clock laser and the frequency comb is continuously monitored by a frequency counter referenced to a 10 MHz hydrogen maser (iMaser 3000), which has a stability of 4$\times$10$^{-16}$ at 10,000 s. As such, the frequency of the probe laser can be calibrated from the clock laser frequency.

The probe beam is split into two branches. One is for the frequency monitor with a Rb oscillator-referenced wavemeter, the other is for the spectroscopy, where counter-propagating configuration is employed to cancel the first-order Doppler shift and the corresponding uncertainties. The entire MOT ensemble is illuminated with a 6 mm-diameter probe beam. The fluorescence is detected by the camera while scanning the probe laser frequency. After each measurement, we optimize experimental parameters to ensure each measurement is performed under the same condition. Each data point of a measurement is an average of five to ten measurements. The RIS is recorded by scanning the probe laser across the resonance.

\section{results and discussion}

\subsection{Absolute frequency measurements of $5s5p^{3}$P$_{0}\to5s6d^{3}$D$_{1}$}

The frequency measurement results and the corresponding uncertainties for all the isotopes are shown in Table~\ref{tab:table1}. The absolute frequencies have been measured with an uncertainty of less than 30 kHz. For $^{87}$Sr, we weight the measurements of three hyperfine manifolds $F\in\{11/2,9/2,7/2\}$ and derive the center of gravity (cog) as well as hyperfine constants $\texttt{A}$ and $\texttt{B}$. We improved the accuracy of hyperfine constants by two orders of magnitude compared to the previous result in Ref. \cite{13}. Due to the separation of 5 cm$^{-1}$ between $^{3}$D$_{1}$ and $^{3}$D$_{2}$, second-order hyperfine interactions are taken into account in the determination of cog of $^{87}$Sr. Details of the theoretical calculations and experimental evaluation for the second-order contribution can be seen in Appendix~\ref{app:secB}. In the determination of isotope shifts, the frequencies of $^{84,86,87}$Sr are referenced to $^{88}$Sr.  

\begin{table*}
\captionsetup{justification=raggedright}
\caption{\label{tab:table1}Absolute frequency measurements of $5s5p^{3}$P$_{0}\to5s6d^{3}$D$_{1}$ for all the Sr isotopes and the isotope shifts relative to $^{88}$Sr. For $^{87}$Sr, the frequency measurements are listed for the first- and second-order perturbation theory. The hyperfine constants $\texttt{A}$ and $\texttt{B}$ are derived in corresponding cases. Results from Refs. \cite{12,13} are also listed for a comparison. The numbers in parentheses indicate 1$\sigma$ uncertainty.}
\begin{ruledtabular}
\begin{tabular}{ccccc}
Isotopes&\multicolumn{2}{c}{Absolute frequency (MHz)}&Previous&Isotope shifts (MHz)\\ \hline
$^{88}$Sr&\multicolumn{2}{c}{760524409.251(25)}&760524989\footnotemark[1]&0\\
$^{84}$Sr&\multicolumn{2}{c}{760524318.199(28)}&&91.052(35)\\
$^{86}$Sr&\multicolumn{2}{c}{760524354.651(26)}&&54.600(33)\\\cline{2-3}&\\
&In 1$^{st}$ order&In 2$^{nd}$ order&\\
$^{87}$Sr, $f_{\texttt{cog}}$&760524336.980(19)&760524357.610(19)&&51.641(28)\footnotemark[3]\\
$^{87}$Sr, F=11/2&760525416.310(31)&760525438.153(31)&\\
$^{87}$Sr, F=9/2&760524087.692(29)&760524111.756(29)&\\
$^{87}$Sr, F=7/2&760523029.704(28)&760523044.114(28)&\\
A&238.984(11)&239.599(11)&239.7(5)\footnotemark[2]\\
B&15.500(34)&9.382(34)&5(20)\footnotemark[2]\\
\end{tabular}
\end{ruledtabular}
\footnotetext[1]{Taken from X. Zhou, $et$ $al$.~\cite{12}.}
\footnotetext[2]{Taken from S. Stellmer and F. Schreck~\cite{13}.}
\footnotetext[3]{Measured result in 2$^{nd}$ order perturbation.}
\end{table*}

\subsection{Shift effects analysis}
We investigate the leading systematic shifts in the frequency measurements. The leading systematic effects and uncertainties are summarized in Table~\ref{tab:table2}. The probe power shift, density shift and misalignment are the leading effects, which have a 70$\sim$200 kHz shift and a 7$\sim$17 kHz uncertainty, while others are less influential. The density-dependent shift is the predominant effect which causes a $\sim$200 kHz shift under the operating condition having a density shift coefficient of -1.4$\times10^{-4}$ Hz cm$^{3}$. The power-induced AC stark shift introduces $\sim$100 kHz contribution at 70 $\mu$W with the coefficient $\kappa=100.98$ kHz mW$^{-1}$ cm$^{2}$. In addition, the probe beam misalignment can shift the resonant frequency, which is verified to be $\sim$100 kHz. The systematic effects and respective uncertainties are analysed in Appendix~\ref{app:secC}.

\subsection{King plot}

Linear King plot can help us understand the relationship of isotope shifts of two distinct transitions, and even access to the knowledge of the origin of isotope shifts for a specific transition. A King plot can be generated by linearly fitting normalized isotope shifts of two transitions by the following expression\cite{46},
\begin{equation}
    \overline{\delta\nu_{i}}^{AA'}=K_{i}-\frac{F_{i}}{F_{j}}K_{j}+\frac{F_{i}}{F_{j}}\overline{\delta\nu_{j}}^{AA'}\label{3}
\end{equation}
Where, $K_{i}$($K_{j}$), $F_{i}$($F_{j}$) denote the electronic mass- and field-shift factors for the transition $i$($j$), respectively; $\overline{\delta\nu_{i}}^{AA'}$($\overline{\delta\nu_{j}}^{AA'}$) is normalized isotope shifts between isotopes $A$ and $A'$ for the transition $i$($j$) given by $\overline{\delta\nu_{i}}^{AA'}$=$\delta\nu_{i}^{AA'}$$\frac{m^{A}m^{A'}}{m^{A}-m^{A'}}$.

\begin{figure}
\centering
\includegraphics[width=8.7 cm]{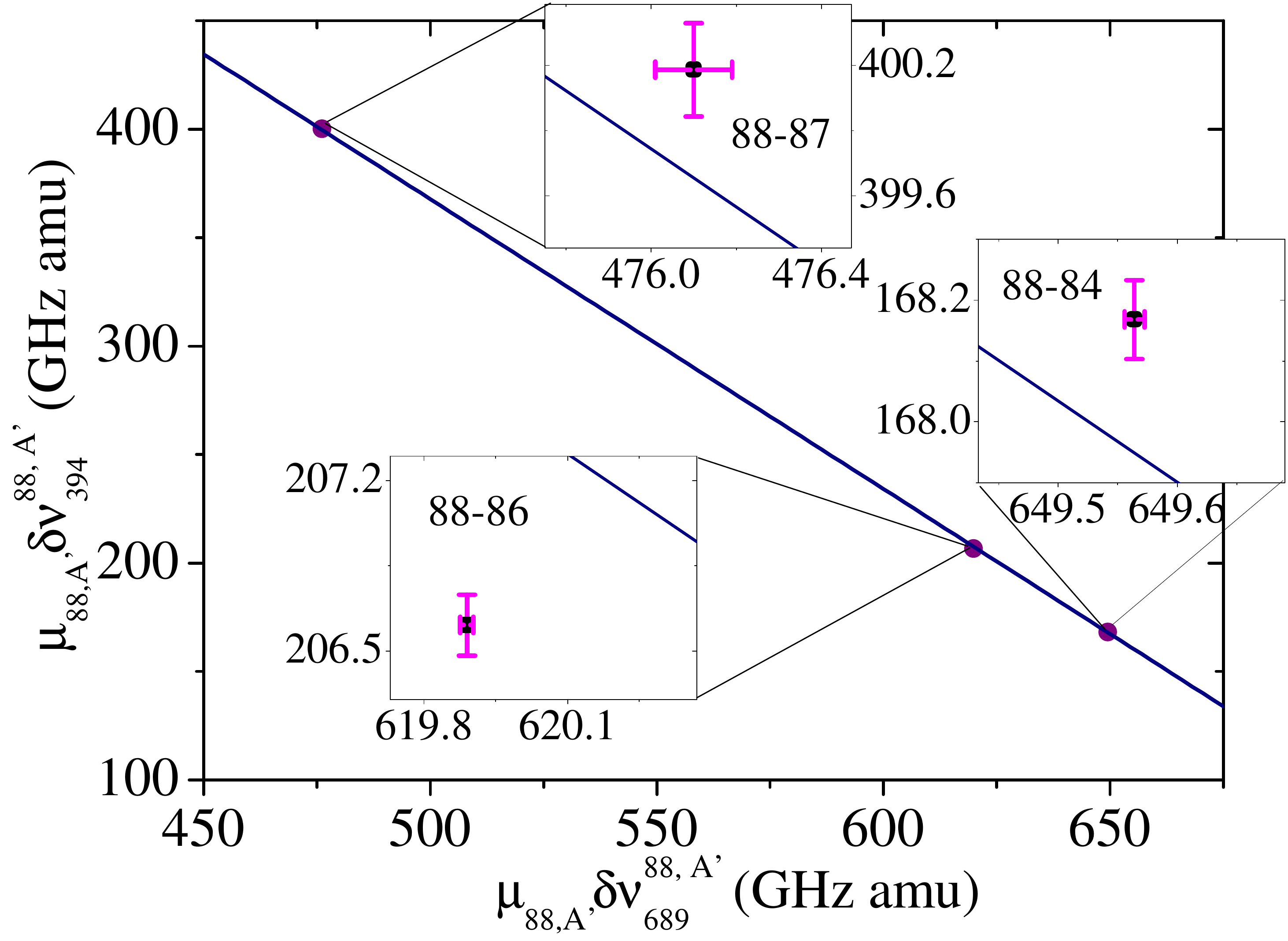}
\captionsetup{justification=raggedright}
\caption{\label{fig:3} 2D King plot of $\gamma$: 394 nm $^{3}$P$_{0}\to^{3}$D$_{1}$ versus $\alpha$: 689 nm $^{1}$S$_{0}$-$^{3}$P$_{1}$. The data for 689 nm is from Ref. \cite{27}. The solid line is a linear fit to the data points. The extracted fit parameters are given in Table~\ref{tab:table4}. The insets show the details at each isotope pair with error bars meaning 1$\sigma$ uncertainty.}
\end{figure}

In order to establish a King plot, we chose $^{3}$P$_{0}$-$^{3}$D$_{1}$ at 394 nm studied in this work as the transition $\gamma$ and $^{1}$S$_{0}$-$^{3}$P$_{1}$ at 689 nm as the transition $\alpha$, which the data is taken from Ref. \cite{27}. The result is shown in Fig.~\ref{fig:3}. The blue straight line is best linear fit to experiment data points. In the fit, the residual sum of squares $\chi^{2}$ is minimized to be 79.06 for one degree of freedom. The fitted line yields a slope of $F_{394}/F_{689}=-1.336\pm0.011$ and an intercept of $K_{394}-F_{394}/F_{689}\cdot K_{689}=1035.71\pm7.20$ GHz amu. With the given values of $\delta\langle r^{2}_{c}\rangle^{84,88}$=0.116(3), $\delta\langle r^{2}_{c}\rangle^{86,88}$=0.050(2)\cite{58}, the field and mass shift constants $F$ and $K$ for the $\gamma$ transition are derived to be 1523.5(5.6) MHz fm$^{-2}$ and -494.9(1.2) GHz amu, respectively. The results are summarized in Table~\ref{tab:table4}. The details of the fitting at each point can be seen by the zoomed insets, showing a deviation of the fitted line to the points. In order to better understand the deviation due to the nonlinearity, we need atomic structure calculations.

\begin{table}
\captionsetup{justification=raggedright}
\caption{\label{tab:table4}Best-fit parameters of King plot in Fig.~\ref{fig:3}. The values of $F_{394,689}$ and $K_{394,689}$ are extracted from the slope and intercept of the linear fitting, respectively. The field shift and mass shift constants $F$ and $K$ are derived from the measured isotope shifts and the given values of $\delta\langle r^{2}_{c}\rangle^{84,88}$=0.116(3) fm$^{2}$, $\delta\langle r^{2}_{c}\rangle^{86,88}$=0.050(2) fm$^{2}$\cite{58}. The units of $K$ and $F$ are given by GHz amu and MHz fm$^{-2}$, respectively.}
\begin{ruledtabular}
\begin{tabular}{cc}
 Parameter&This work\\ \hline
$F_{394,689}$&-1.336(11)\\
$K_{394,689}$&1035.7(7.2)\\
$F_{394}$&1523.5(5.6)\\
$K_{394}$&-494.9(1.2)
\end{tabular}
\end{ruledtabular}
\end{table}

\section{CONCLUSIONS}
We have performed a repumping induced spectroscopy for the $5s5p^{3}$P$_{0}\to5s6d^{3}$D$_{1}$ transition of Sr in a MOT. We have measured the absolute frequencies of the transition for all the stable isotopes with an accuracy of less than 30 kHz, by taking the primary shift effects into account. In addition, we have calculated the hyperfine constants A and B for the fermionic isotope. Combining the data from the reference \cite{27} for the 689 nm transition, we have performed a 2D King plot. Our measurements may help the search for theoretically predicted new light bosons with alkaline-earth-metal atoms.

\begin{acknowledgments}
We thank Vladimir A. Dzuba for insightful discussions. We thank Ceren Yuce for helpful comments on the final manuscript revision. We thank Qiushuo Sun for her contribution on the stabilization of the clock laser. This work is supported by European Union's Horizon 2020 Research and Innovation Programme under the Marie Sklodowska-Curie grant agreement No. 860579 (MoSaiQC Project) and No. 820404 (iqClock project) and ICON (International Clock and Oscillator Networking): EP/W003279/1. 

\end{acknowledgments}

\begin{figure*}[htbp]
\centering
\begin{subfigure}[t]{
   0.3\linewidth}
\centering
\includegraphics[width=5.1 cm]{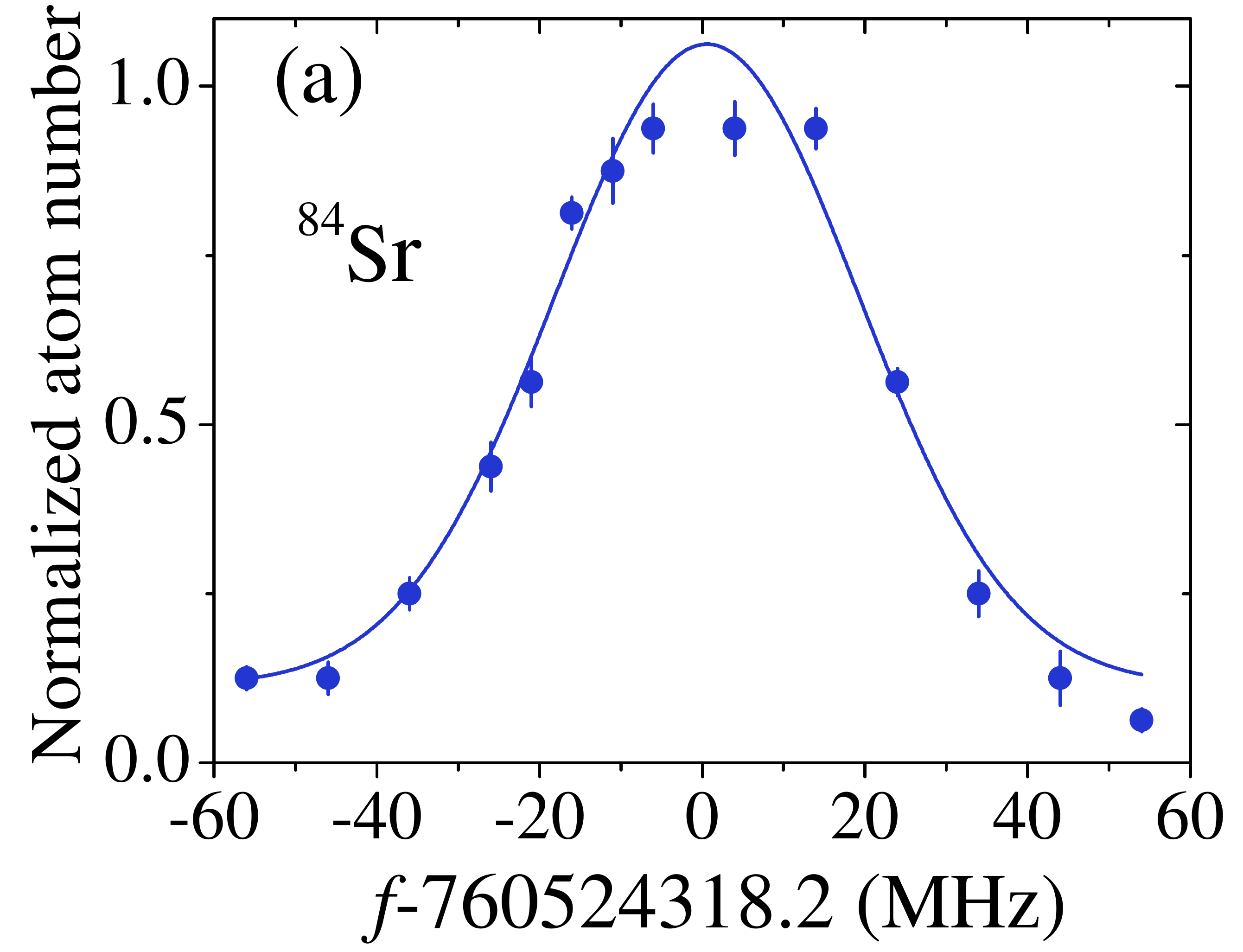}
\end{subfigure}
\begin{subfigure}[t]{
    0.3\linewidth}
\centering
\includegraphics[width=4.5 cm]{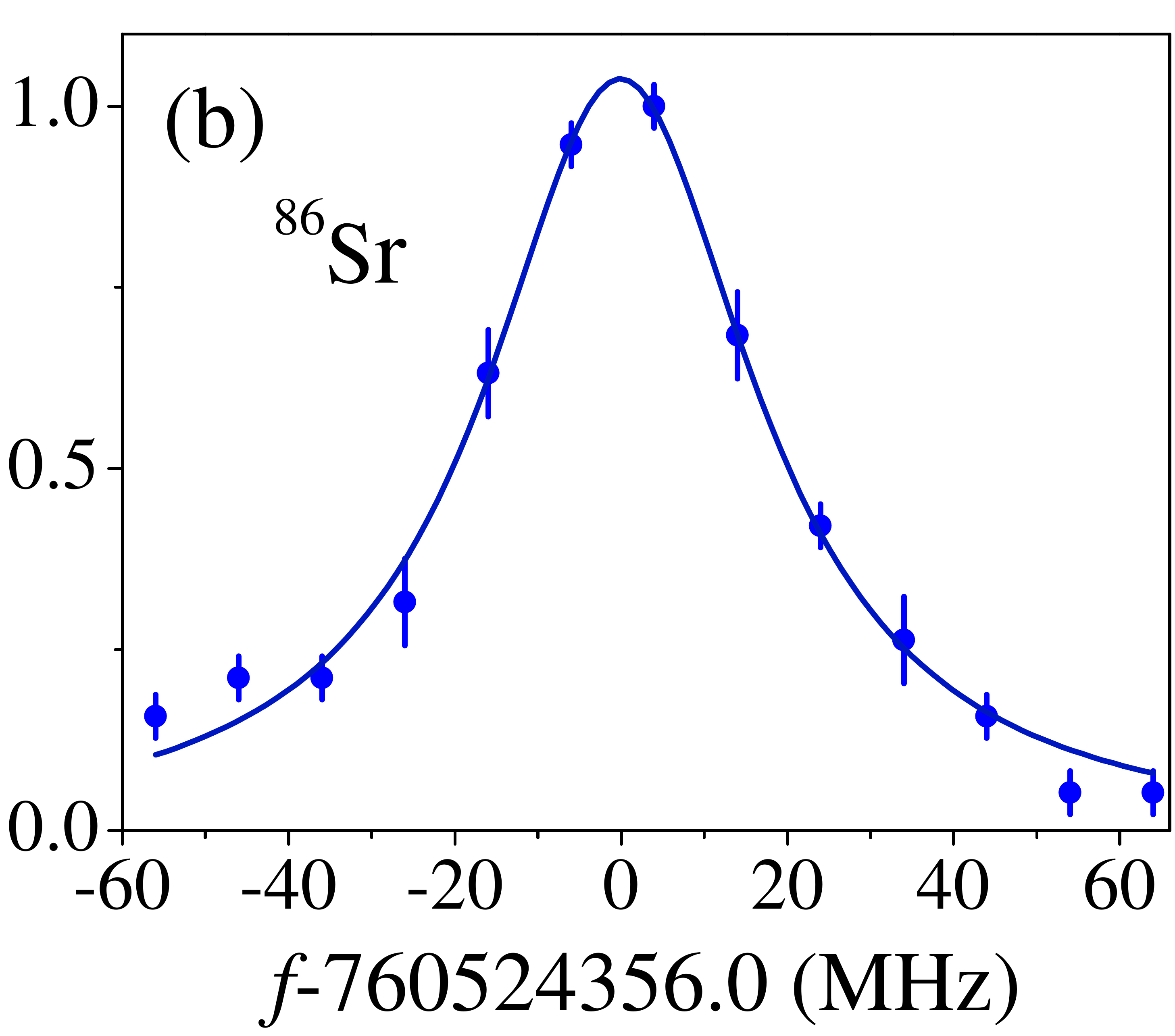}
\end{subfigure}
\begin{subfigure}[t]{
    0.3\linewidth}
\centering
\includegraphics[width=4.8 cm]{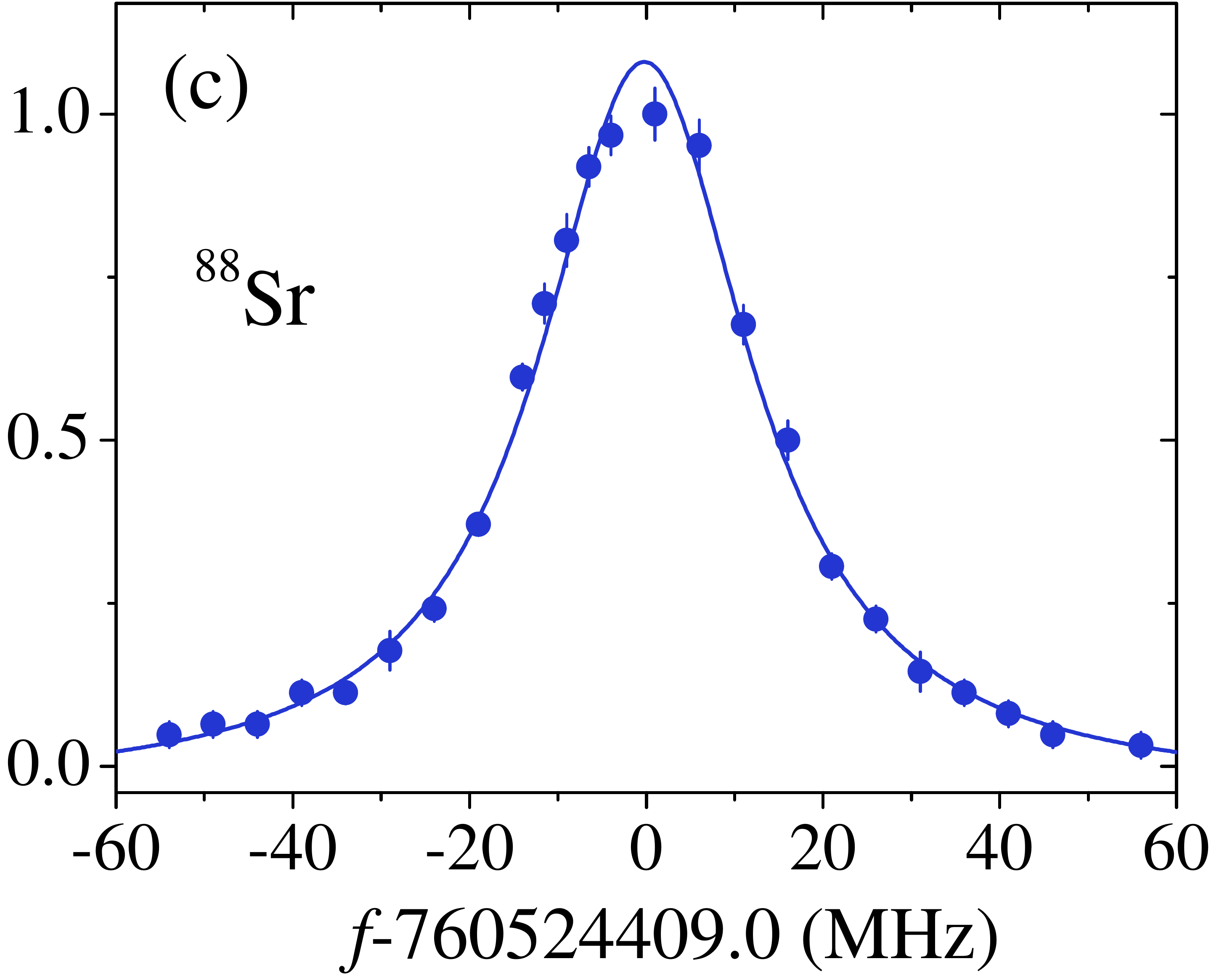}
\end{subfigure}
\quad
\begin{subfigure}[t]{
    0.3\linewidth}
\centering
\includegraphics[width=4.9 cm]{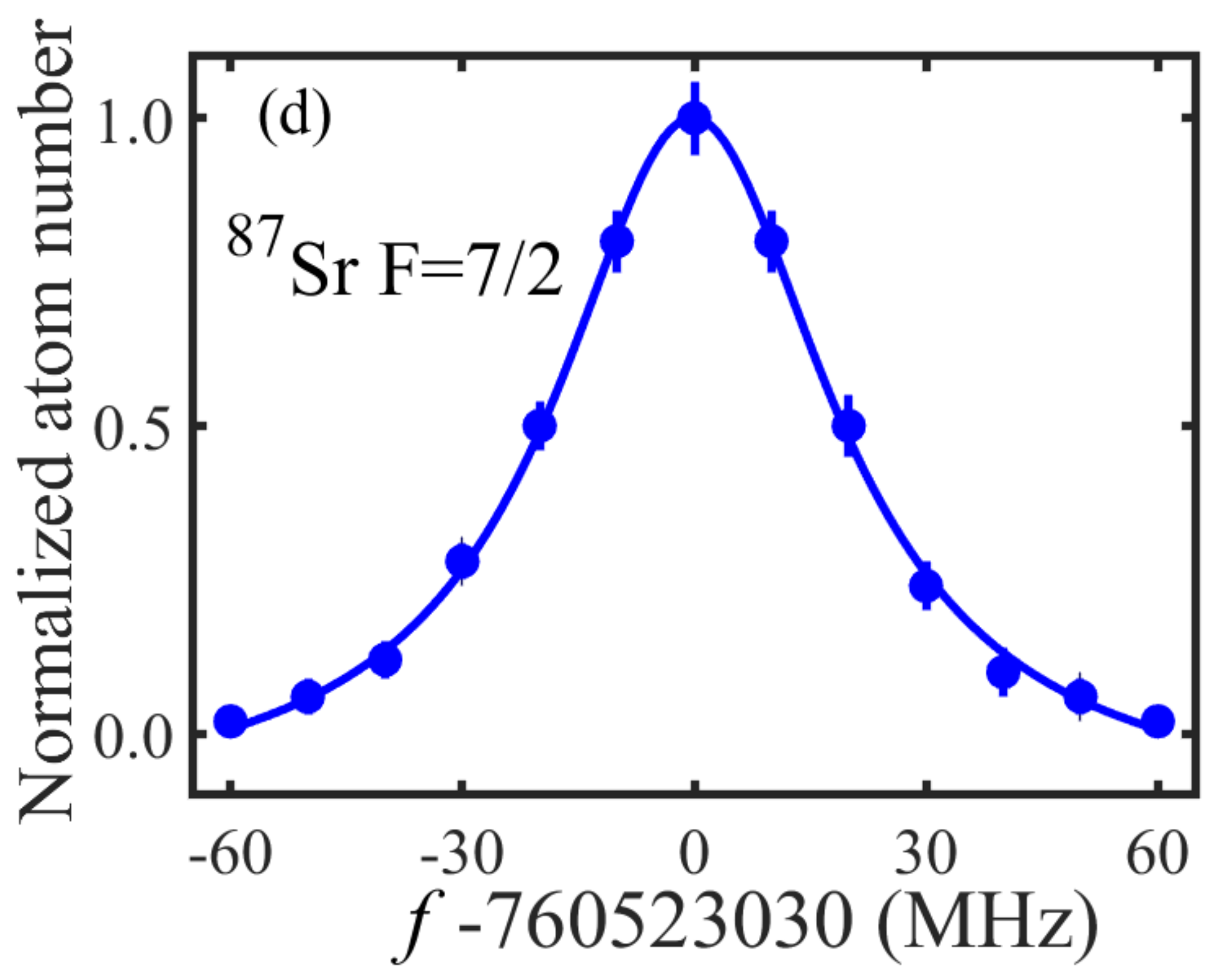}
\end{subfigure}
\begin{subfigure}[t]{
    0.3\linewidth}
\centering
\includegraphics[width=4.9 cm]{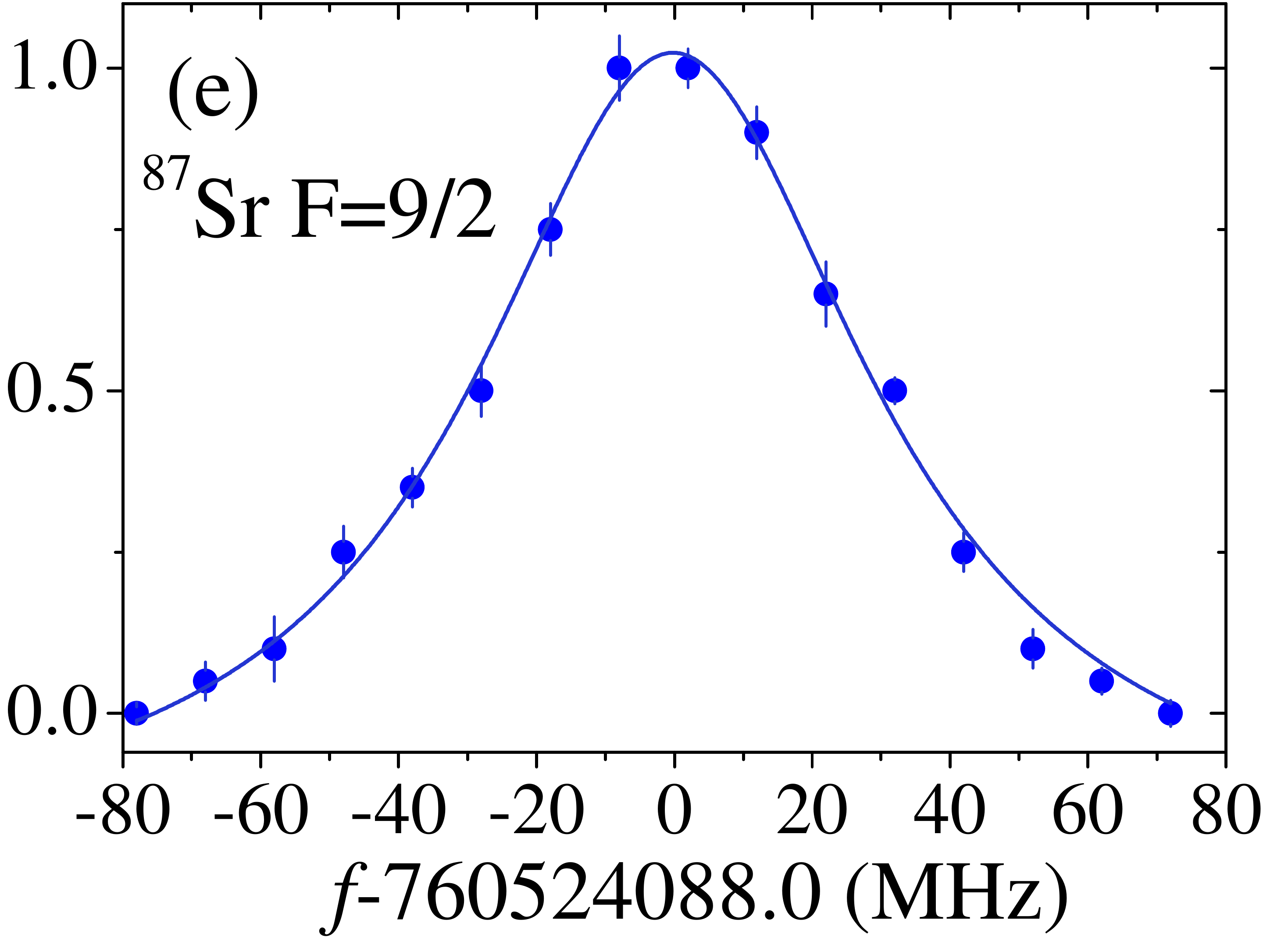}
\end{subfigure}
\begin{subfigure}[t]{
    0.3\linewidth}
\centering
\includegraphics[width=4.8 cm]{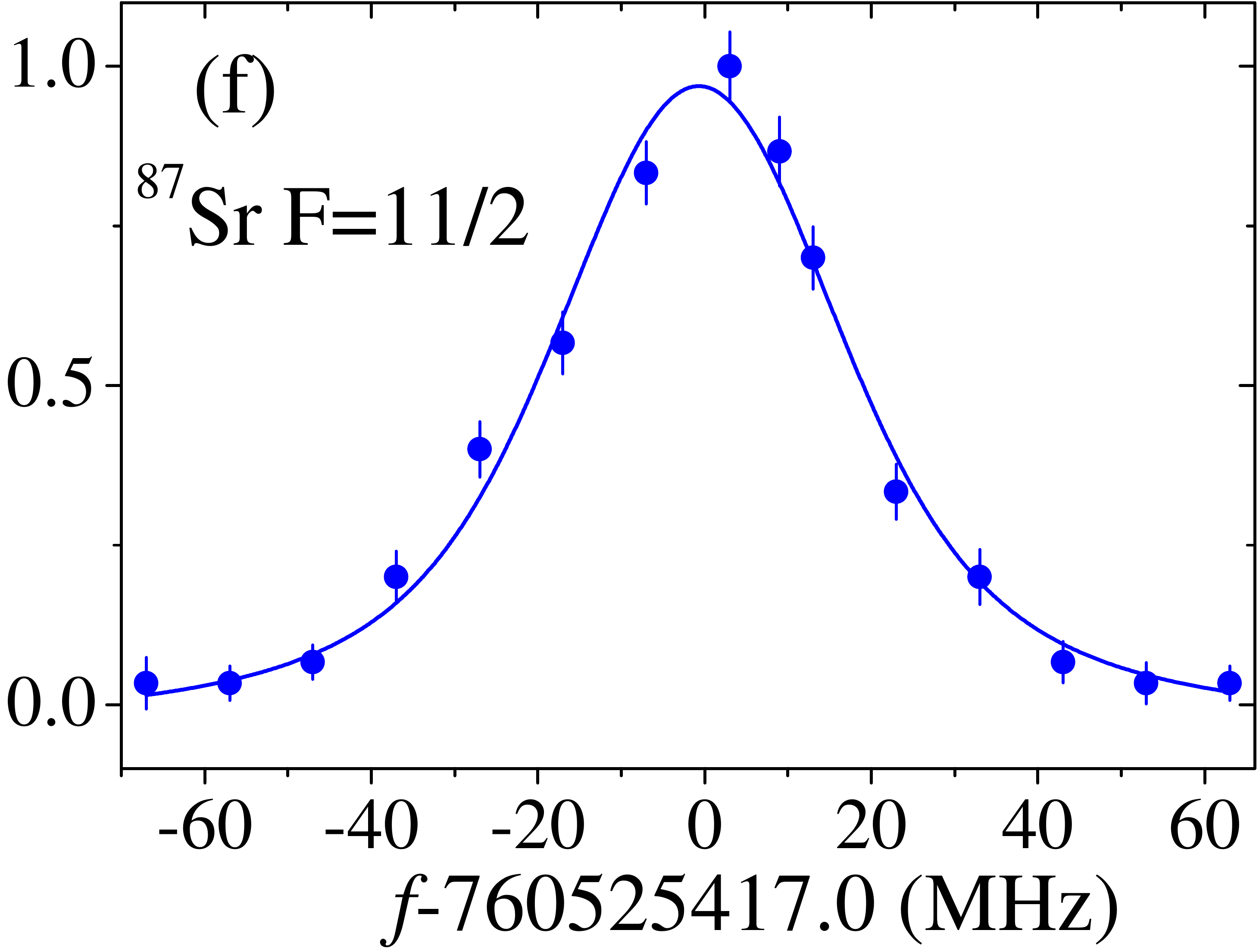}
\end{subfigure}
\captionsetup{justification=raggedright}
\caption{\label{fig:7} RIS of $^{3}$P$_{0}\to^{3}$D$_{1}$ for all the Sr isotopes. All the measurements are performed at 70 $\mu$W. The measurement for $^{84}$Sr is carried out at higher oven temperature and higher magnetic field in order to increase the atom number in the ensemble, which results in broader spectroscopic linewidth. Each data point is the average of 5-10 measurements. The atom number is normalized. The solid line is a Lorentzian fit to data. The error bars indicate 1$\sigma$ statistical uncertainty.}
\end{figure*}

\appendix{\section{\uppercase{Repumping induced Spectroscopy}}}
Repumping induced spectroscopy for all the stable isotopes is shown in Fig.~\ref{fig:7}. The probe beam is in a retro-reflection configuration to cancel the Doppler effect and the power is set at 70 $\mu$W to reduce the power-induced frequency shift and broadening. 
\begin{table*}
\captionsetup{justification=raggedright}
\caption{\label{tab:table2}Systematic frequency shifts and uncertainties (kHz) for the $5s5p^{3}$P$_{0}\to5s6d^{3}$D$_{1}$ transition for all the Sr isotopes. Uncertainties indicate 1$\sigma$ deviation.}
\begin{ruledtabular}
\begin{tabular}{ccccccccc}
 &\multicolumn{2}{c}{88}&\multicolumn{2}{c}{84}&\multicolumn{2}{c}{86}&\multicolumn{2}{c}{87}\\
 Contributors&Corr. (kHz)&Unc. (kHz)&Corr. (kHz)&Unc. (kHz)
&Corr. (kHz)&Unc. (kHz)&Corr. (kHz)&Unc. (kHz)\\ \hline
 Probe power shift&-94&13&-112&16&-101&14&-73&8 \\
 Density shift&183&16&203&17&197&16&142&11\\
 Recoil shift&-15&$<$0.1&-15&$<$0.1&-15&$<$0.1&-15&$<$0.1\\
 Misalignment&93&9.3&101&10.1&96&9.6&71&7.1\\
 2$^{nd}$ order Doppler shift&2$\times$10$^{-6}$ &$<$0.1$\times$10$^{-6}$&2$\times$10$^{-6}$&$<$0.1$\times$10$^{-6}$&2$\times$10$^{-6}$&$<$0.1$\times$10$^{-6}$&2$\times$10$^{-6}$&$<$0.1$\times$10$^{-6}$\\
 Line profile&0&4&0&4&0&4&0&4\\
 Quadratic Zeeman shift&5.5&$<$0.1&5.5&$<$0.1&5.5&$<$0.1&4.3&$<$0.1\\
 Laser frequency calibration&0&10&0&10&0&10&0&10\\
 Total&173&25&183&28&183&26&129&19
\end{tabular}
\end{ruledtabular}
\end{table*}
To obtain the spectroscopy for each isotope, the atoms are loaded into a blue MOT and the probe laser is scanned across the resonance. We keep all experimental parameters the same for all isotopes except the oven temperature for $^{84}$Sr. We slightly increase the oven temperature to introduce higher flux for $^{84}$Sr due to the low abundance of 0.56\%. To improve the signal to noise ratio of spectroscopy, the measurement at each frequency point is repeated 5 to 10 times and the mean value is chosen as the data point. The resolved hyperfine structure of $^{3}$D$_{1}$ in $^{87}$Sr is shown in Fig.~\ref{fig:7}(d)-(f). The experimental data are fitted with a Lorentzian function and the results show that the average linewidths are in a range of 30 to 50 MHz. The linewidth broadening is estimated to be dominated by the magnetic field during the MOT operation. Note that the distorted or asymmetric characteristics seen in some spectroscopic examples, such as $^{84}$Sr and $^{86}$Sr, can be accounted for by the lower atom number in the MOT. With the same experimental condition, the peak atom number in blue MOT of $^{88}$Sr is an order of magnitude higher than $^{86}$Sr and $^{87}$Sr, and two orders of magnitude higher than $^{84}$Sr. The atom fluctuation can be 10\%$\sim$20\% during the measurement process for each isotope. To eventually determine the resonance frequency for each isotope, we perform multiple spectroscopic measurements and take the averaged value as the absolute frequency. The systematic frequency shifts and corresponding uncertainties are summarized in Table~\ref{tab:table2}.

\section{\uppercase{Characterization of different freqeuncy shift effects}\label{app:secC}}
\subsection{\label{app:subsec}Density shift}
\begin{figure}
\includegraphics[width=10.4 cm]{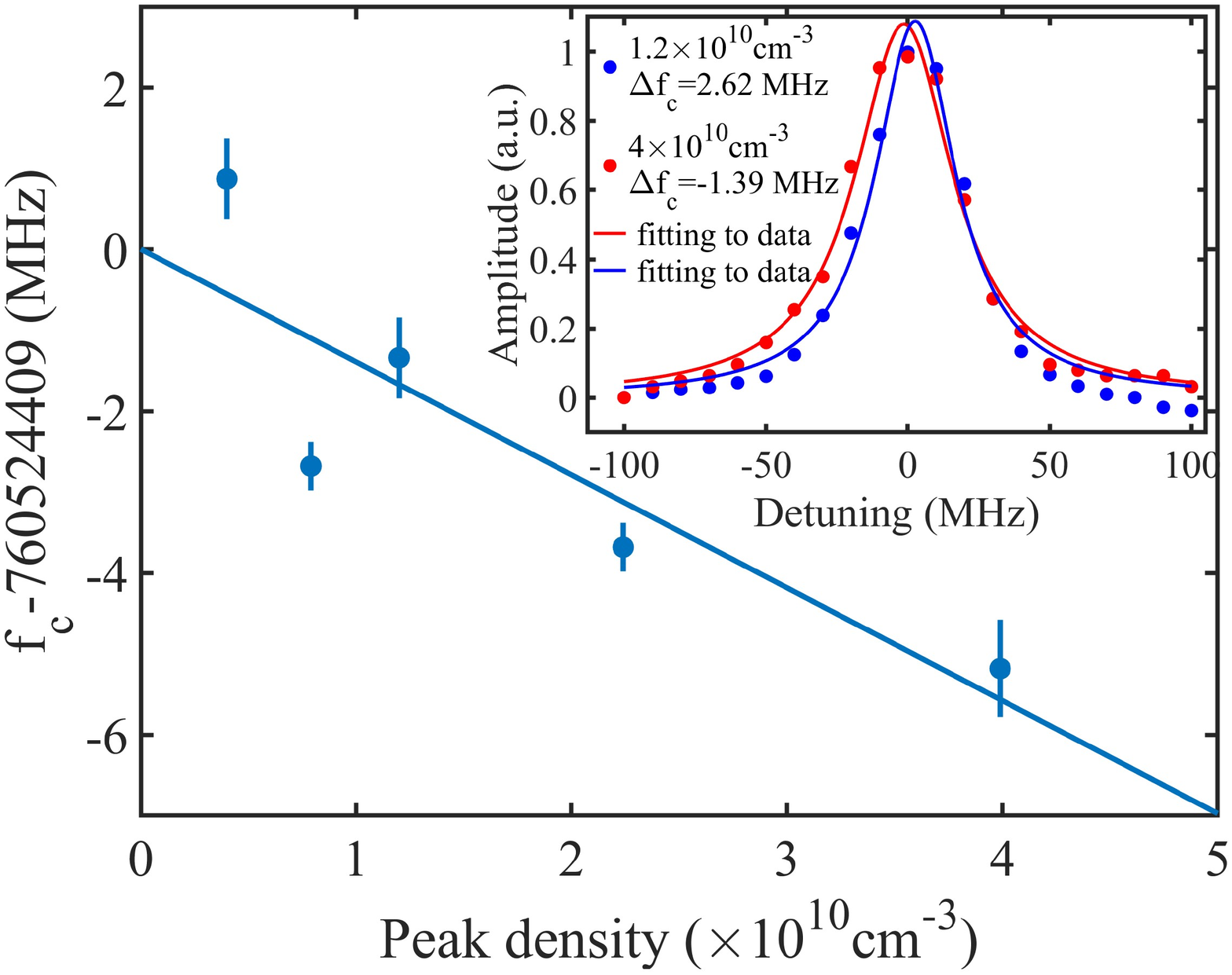}
\captionsetup{justification=raggedright}
\caption{\label{fig:8}Dependence of line centre of $^{3}$P$_{0}\to^{3}$D$_{1}$ on the peak density from 4$\times10^{9}$ to 4$\times10^{10}$ cm$^{-3}$ for $^{88}$Sr. The filled circles and error bars are data and 1$\sigma$ uncertainty, respectively. The red line is a fit. The density shift coefficient is -1.4$\times10^{-4}$ Hz cm$^3$. The inset shows the spectra with a Lorentzian fit for densities of 1.2$\times$ and 4$\times10^{10}$ cm$^{-3}$.}
\end{figure}

The density dependent frequency shift \cite{6,7} is one of the leading systematic effects in our experiment. We experimentally determine this effect by performing spectroscopic measurements at various atomic densities ranging from 10$^9$ to $4\times10^{11}$ cm$^{-3}$. The density is varied by changing the oven temperature or the magnetic field gradient. The maximum density we can reach is 4$\times10^{11}$ cm$^{-3}$ by optimizing the magnetic field at $\sim$65 G/cm and the oven temperature at 1063 K. For the measurements at each density, the operating probe power is set at 70 $\mu$W to minimize the probe power-induced AC stark shift. The ensemble is operated at a density of 2$\times10^{9}$ cm$^{-3}$. To extract the density shift, we apply a linear fit to the experimental data and extrapolate to zero density. Fig.~\ref{fig:8} illustrates a typical example of the density-dependent frequency shift as a function of the density from 4$\times10^{9}$ to 4$\times10^{10}$ cm$^{-3}$ for $^{88}$Sr. Fitting the measured data points by a straight line reveals the density shift coefficient of -1.4$\times10^{-4}$ Hz cm$^3$. Note that the far deviation of data points at low densities from the fitted line is due to the low signal-to-noise ratio of fluorescence spectroscopy at low densities resulting in an increased bias in determining the resonant frequency. The inset shows spectra at 1.2$\times$ and 4$\times10^{10}$ cm$^{-3}$ density and fits with Lorentzian function, which depicts the resonance frequency to be red detuned by 3.8 MHz with linewidth broadening from 1.2$\times10^{10}$ to 4$\times10^{10}$ cm$^{-3}$. Note that the frequency measurement for $^{84}$Sr is performed only density is of 10$^9\sim10^{10}$ cm$^{-3}$ due to its low abundance.

\subsection{\label{app:subsec}Probe power shift}
\begin{figure}
\centering
\includegraphics[width=8.6 cm]{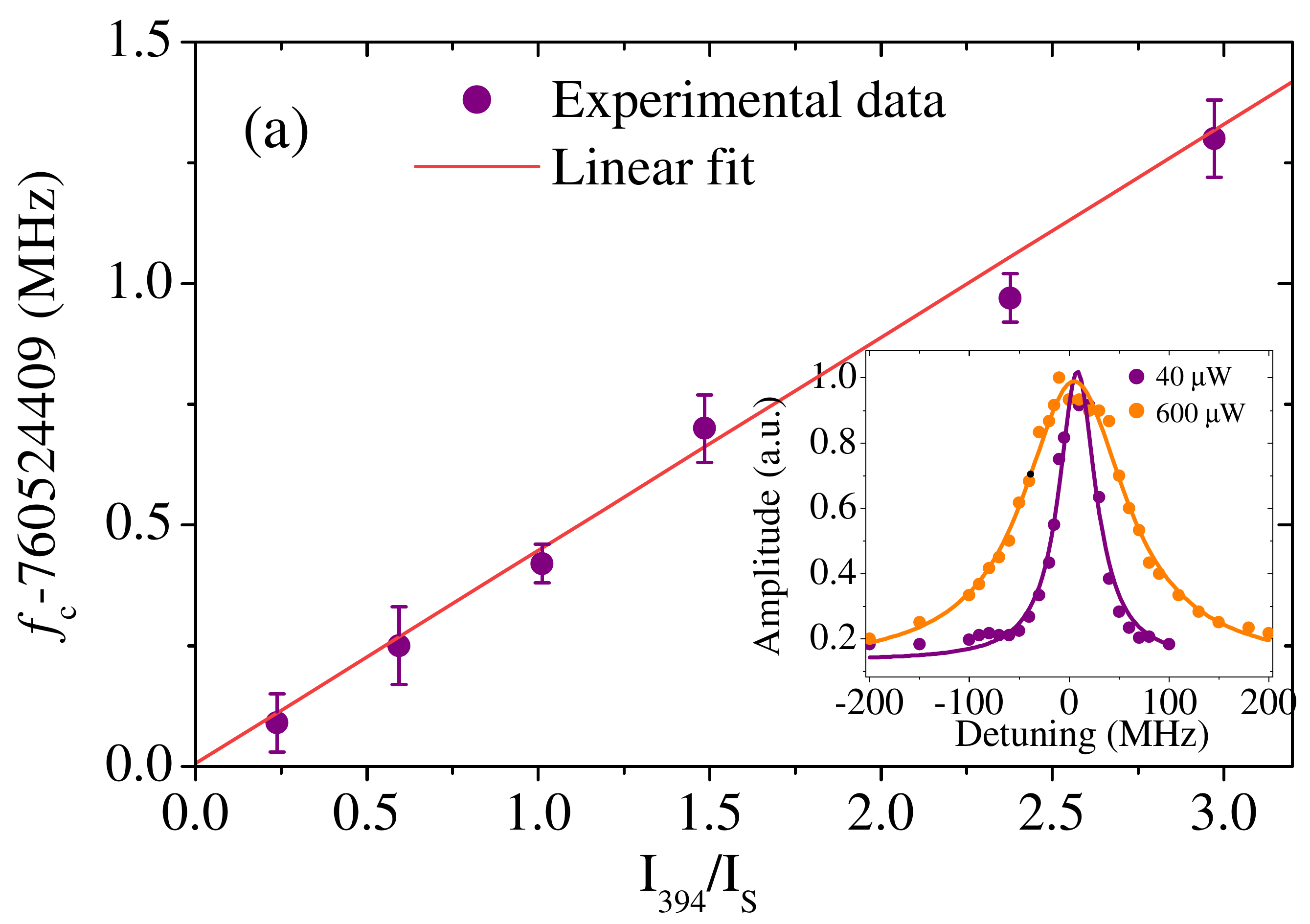}
\quad
\includegraphics[width=8.6 cm]{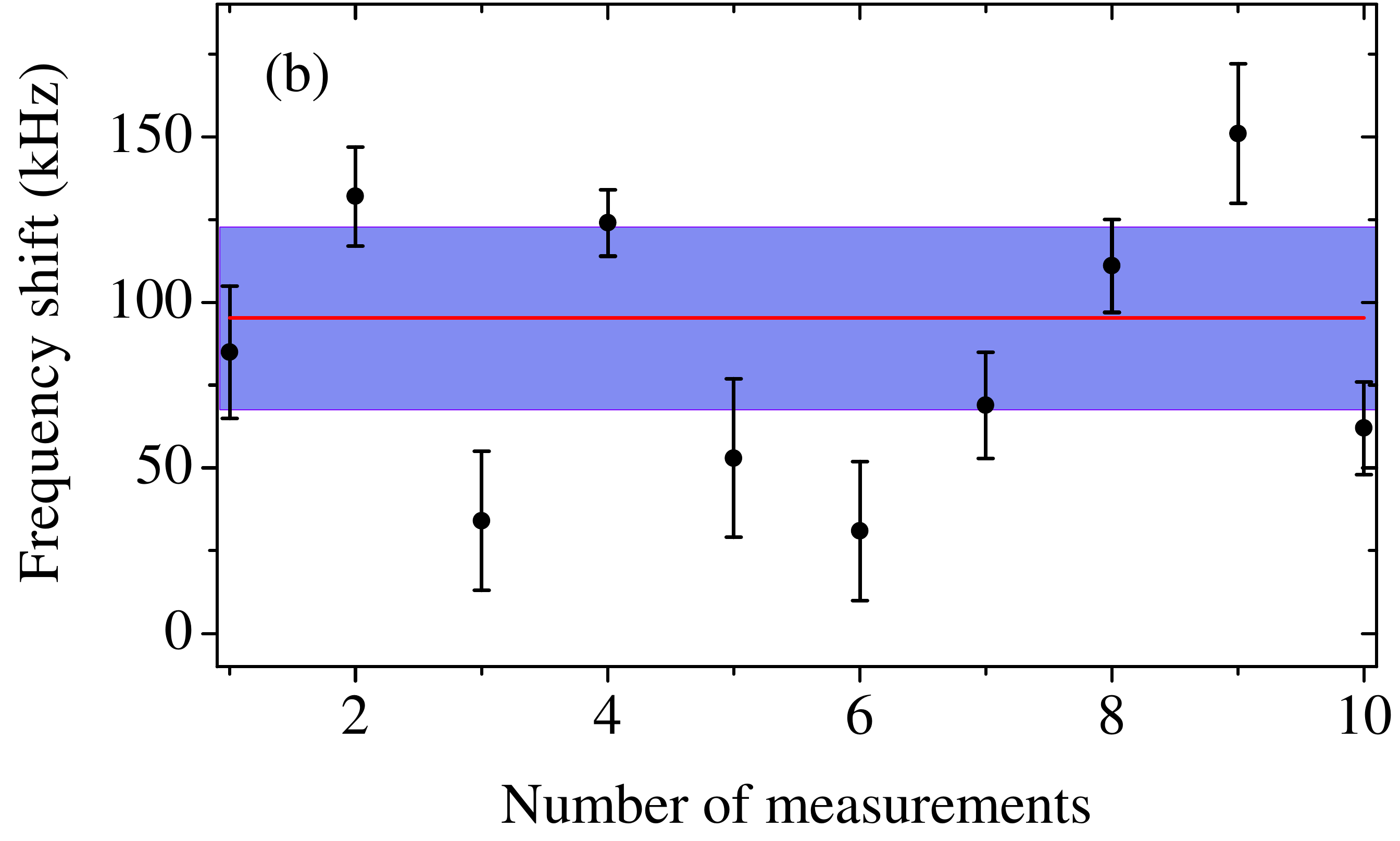}
\captionsetup{justification=raggedright}
\caption{\label{fig:9} (a) Dependence of line centre of $^{3}$P$_{0}\to^{3}$D$_{1}$ on the probe intensity for $^{88}$Sr. The filled circles and error bars are experimental data and 1$\sigma$ uncertainty, respectively. The red line is a fit to the data points. The inset shows the spectra for the probe power of 40 and 600 $\mu$W. (b) Ten-times repeated measurements of $^{88}$Sr frequency. The red line is a weighted mean value of 94 kHz and the shaded area indicates the 95$\%$ confidence interval (CI) of 55.1 kHz.}
\end{figure}
Another leading systematic contributor is the probe power-induced AC stark shift \cite{4} arising from differential polarizabilities of $^3$P$_0$ and $^3$D$_1$ states. To evaluate this shift $\kappa$I (where $\kappa$ is the shift coefficient, I is the probe intensity), we drive the $^3$P$_0$-$^3$D$_1$ transition for various intensities of the probe laser in a range of 0.2$\sim$3I$_S$ (where I$_S$=4.76 mW/cm$^{2}$ being the saturation intensity of this transition). We extrapolate the linear fit to zero intensity for all the isotopes. Fig.~\ref{fig:9} shows the power-induced frequency shift. At each point, we scan the probe frequency up and down crossing the resonance to average hysteresis effects. An example of the probe power shift for $^{88}$Sr is illustrated in Fig.~\ref{fig:9}(a). The red line is a fit to the data points, revealing the power shift efficient $\kappa$=100.98 kHz mW$^{-1}$ cm$^2$. At the operating power of 70 $\mu$W the shift is $\sim$100 kHz with respect to zero point. The inset shows the spectra for two different powers, which indicates the frequency is shifted by 3.1 MHz for 600 $\mu$W relative to 40 $\mu$W. The same measurement procedure is repeated 10 times for each isotope. After each measurement, the setup including laser polarization and beam alignment is optimized to ensure all data are recorded under the same condition. Fig.~\ref{fig:9}(b) shows the mean value of the frequency shift to be 94 kHz and the 1$\sigma$ uncertainty to be 13 kHz for $^{88}$Sr. 


\subsection{\label{app:subsec}Misalignment and line profile}
Due to the imperfect wavefront overlapping \cite{11}, the misalignment of the retro-reflected probe beam is verified to be a key factor in our experiment for the frequency shifts and uncertainties. To assess this effect, we follow the proposed method in Ref \cite{3} by deliberately misaligning the retro-reflected beam until the resonance intensity is reduced to 50$\%$. A frequency shift in a range of 0.2$\sim$1 MHz is observed for each isotope. The frequency shift due to the misalignment is estimated to be 40$\sim$200 kHz. The resulting uncertainty is determined to be 10$\%$ of the frequency shift. In addition, the probe beam may cause the asymmetry in the recorded spectral line profile \cite{5} due to the retro-reflected power loss by 5$\sim$8$\%$ due to the view port and the mirror. In fact, even though a significant shift by the line profile asymmetry is not experimentally observed for 70 $\mu$W of the probe laser, the uncertainty is estimated to be 4 kHz for each isotope.

\subsection{\label{app:subsec}Quadratic Zeeman shift}
Since $\pi$ transitions ($\Delta m$=0) are driven by a linearly polarized probe beam, the linear Zeeman shift cancels out for all the isotopes. However, quadratic Zeeman shift \cite{8,32} arising from Zeeman coupling between $^3$D$_1$ and $^3$D$_2$ states is a considerable factor when the remaining magnetic field is non-negligible in our experiment. As the separation of $^{3}$P$_{\texttt{J}}$ states is much larger than that of $^{3}$D$_{\texttt{J}}$, we only consider the quadratic Zeeman shift of $^3$D$_1$. To estimate the quadratic Zeeman shift, we calculate Zeeman matrix elements $<^3$D$_1|H_z|^3$D$_2>$ in LS coupling for all the isotopes. For even isotopes, the matrix element $<^3$D$_1, m=0|H_z|^3$D$_2, m=0>$ is calculated to be 8.2 MHz mT$^{-1}$B, yielding a quadratic Zeeman shift of 5.5 kHz at the magnetic field of $\sim$3.5 mT introduced by surrounding ion pumps and Zeeman slower magnets. Yet in terms of the odd isotope, i.e., $^{87}$Sr, each Zeeman substate is coupled with multiple substates of $^3$D$_2$, which dramatically increases the computation complexity. In this work, we focus on $|F,m=11/2\rangle$, $|F,m=9/2\rangle$ and $|F,m=7/2\rangle$ substates of $^{3}$D$_{1}$ to simplify the computing. The pure LS matrix elements are given in Ref\cite{2}.

For $^{3}$D$_{1}$ $|F,m=11/2\rangle$, the matrix elements $|\langle^3$D$_1,F,m=11/2|H_z |^3$D$_2,F,m=11/2\rangle|^2$, $|\langle^{3}$D$_{1},F,m=11/2|H_z|^{3}$D$_{2},F=13/2,m=11/2\rangle|^{2}$ are calculated to be 34, 15 MHz$^{2}$mT$^{-2}$B$^{2}$, respectively. For the $\Delta m$=0 transition, the quadratic Zeeman shift $\Delta\nu_{Z2}$ is given by \cite{1}
\begin{equation}
\Delta\nu_{Z2}=\sum_{F'}\frac{|\langle^3D_1,F,m_F|H_Z|^3D_2,F',m_F\rangle|^2}{\nu_{^3D_2,F'}-\nu_{^3D_1}}\label{eq:C3}
\end{equation}
Thus, by summing up hyperfine states $F'$ of $^{3}$D$_{2}$ with the same $m_{F}$, the quadratic Zeeman shift can be calculated. The shift of $|F,m=11/2\rangle$ is calculated by Eq.~(\ref{eq:C3}) to be 4.00(4) kHz at the magnetic field of 3.5 mT. Similarly, other matrix elements $|\langle^{3}$D$_{1},F,m=9/2|H_z|^{3}$D$_{2},F,m=9/2\rangle|^{2}$, $|\langle^{3}$D$_{1},F,m=9/2|H_z|^{3}$D$_{2},F=11/2,m=9/2\rangle|^{2}$, $|\langle^{3}$D$_{1},F,m=7/2|H_z|^{3}$D$_{2},F,m=7/2\rangle|^{2}$ and $|\langle^{3}$D$_{1},F,m=7/2|H_z|^{3}$D$_{2},F=9/2,m=7/2\rangle|^2$ are also calculated to be 52, 11, 47 and 5 MHz$^{2}$mT$^{-2}$B$^{2}$, respectively. The resulting quadratic Zeeman shifts for $|F,m=9/2\rangle$ and $|F,m=7/2\rangle$ are 5.00(5) kHz and 4.00(4) kHz, respectively, where the numbers in the bracket are the uncertainty due to the measurement of the magnetic field by 10$\%$ error. Therefore, by weighting the shifts of three Zeeman substates, the quadratic Zeeman shift of the centre of gravity for $^{87}$Sr is calculated to be 4.30(4) kHz. 

\subsection{\label{app:subsec}Other shifts}
We calculate two other freqency shifts, i.e., photon recoil shift and the second-order Doppler shift. The photon recoil frequency shift is calculated to be 15 kHz by $\delta\nu=h/(2m\lambda^2)$ \cite{5}, and the second-order Doppler shift is 2 mHz according to $\delta\nu=v^2f/(2c^2)$ \cite{10} at the most probable speed of 0.7 m/s of the ensemble during the measurements. Other shifts are not considered as they are negligible in our measurements.

\section{\uppercase{Transfer cavity locking}}
There exist two reasons that we apply the transfer cavity locking scheme to the probe laser. First, there doesn't exist a frequency comb in our lab to directly measure the frequency of 394 nm laser; second, locking the probe laser to the wavemeter with an accuracy of 2 MHz limits the accuracy of measurements. Therefore, to tackle the aforementioned issue, we link the probe laser to our ultrastable clock laser at 698 nm which can be measured in frequency with our frequency comb. Through this scheme, the accuracy of the probe laser frequency can be determined by that of the clock laser. 394 nm light is generated from 788 nm by frequency doubling it in a Ti:sapphire laser. Specifically, we lock the 788 nm laser to a high-finesse cavity, through a transfer cavity, and derive the probe laser frequency. While this is an indirect frequency evaluation for the probe laser, it can dramatically improve the accuracy of the probe laser frequency by orders of magnitude with respect to measurements by wavemeter.

In a transfer cavity locking scheme, the stability of a master laser is transferred to a slave laser by locking the slave laser to the cavity referenced to the master laser. The experimental implementation is described as follows. The superimposed beams of 788 nm (slave) and 698 nm (master) are coupled into the transfer cavity (FPI 100-0750-3V0, Toptica), the transmission signal are detected by a photodiode (PD) and input to a Python-programmed microcontroller (Red Pitaya STEMlab 125-14). While scanning the cavity over one free spectral range (FSR) of 1 GHz, the transmission peaks for two beams are detected simultaneously. From the start of trigger, the peak position is set in an order of master-slave-mater. By executing the algorithm for peak detection, an error signal is generated and fed back to the cavity controller as well as the slave laser controller to lock them to the master laser. 

The slave frequency $f_{s}$ can be expressed in terms of the master frequency $f_{m}$ \cite{17},
\begin{equation}
    f_{s}=\frac{N_{s}}{N_{m}}(\Delta_{\texttt{FSR}}r+f_{m})\label{eq.a1}
\end{equation}
where $N_{m}$, $N_{s}$ are the master and slave mode numbers, respectively, $r=\frac{t_{m}-t_{s}}{t_{m'}-t_{m}}$ is the ratio of the timing interval between master and slave peaks to FSR. Mode numbers can be estimated from the wavemeter measurement, where $f_{m}$ is measured by the frequency comb, given by
\begin{equation}
    f_{m}=Nf_{\texttt{rr}}+ f_{\texttt{ceo}}-f_{\texttt{beat}}\label{eq.a2}
\end{equation}
where $f_{\texttt{ceo}}=10$ MHz, $f_{\texttt{rr}}=125$ MHz. By substituting $f_{m}$ in Eq.~(\ref{eq.a1}) with Eq.~(\ref{eq.a2}), the probe laser frequency can, therefore, be derived.

The frequency resolution $\Delta f$ of the locking system is limited by the clock speed $\Delta t$ of the microcontroller and the scanning rate $f_{sr}$, given by\cite{31}
\begin{equation}
    \Delta f=2\Delta_{\texttt{FSR}}f_{sr}\Delta t
\end{equation}
where $\Delta t$=10 ns, $f_{sr}=$1 kHz. Hence, this allows the locking system to posses the frequency resolution of 10 kHz, which is good enough for our isotope shift measurements. 

The clock laser is PDH-stabilized to a vertical cavity (finesse=2.3$\times10^{5}$, cavity length=78 mm). By beating with another clock laser system (MenloSystems ORS Ultrastable lasers), we obtain the linewidth of the clock laser to be $\sim$100 Hz. The 1 s fractional stability is 1.1$\times10^{-14}$, corresponding to 4.7 Hz. The drift of Allan deviation is attributed to the temperature drift of the cavity length. The stability of the stabilized slave laser is improved within the whole range with respect to the free-running laser, in particular stability at 700 s improved by a factor of 6000, which confirms that the long term stability of the clock laser is transferred to the 788 nm laser. The best achievable stability is limited by the response speed of PZT of the cavity \cite{1}. The laser noise is able to be probed only at frequencies less than half of the scanning rate and the feedback signal is effective only at frequencies below 1/4 of the scanning rate. We have a maximum scanning rate of 1 kHz which limits the locking loop bandwidth to 250 Hz. 

\section{\uppercase{Second-order contribution}\label{app:secB}}
\subsection{\label{app:sec81}Theoretical calculations}
In the evaluation of hyperfine splittings, it is sufficient, in most situations to consider only the first-order perturbation theory. However, when the level separation is considerably small, e.g., $<$10 cm$^{-1}$, the second-order contribution has to be included, especially in the King plot analysis \cite{30}. From the NIST spectra database \cite{14}, it is known that the separation between 5$s$6$d^3$D$_1$ and $^3$D$_2$ is 5 cm$^{-1}$, which means the second-order effect cannot be neglected here. In this section, we introduce the calculation of second-order hyperfine splittings for all three hyperfine folds and the correction of the center of gravity as well as hyperfine constants. In addition, we evaluate the second-order shift from the experimental point of view by 2D King plot.

The second-order hyperfine splitting of the state $\vert\gamma JIF\rangle$ is given by
\begin{equation}
   \Delta E^{(2)}_{F}=\sum_{\gamma'J'\neq\gamma J}\frac{|\langle\gamma JIF|H_{\texttt{hfs}}|\gamma'J'IF\rangle|^2}{E_{\gamma J}-E_{\gamma'J'}}
\end{equation}
where $\vert\gamma JIF\rangle$ and $\vert\gamma' J'IF\rangle$ represent $\vert^{3}$D$_{1},F\rangle$ and $\vert^{3}$D$_{2},F\rangle$, respectively. The Hamiltonian $H_{\texttt{hfs}}=H_{\mu}+H_{Q}$ consists of magnetic dipole interactions $H_{\mu}$ and electric quadrupole interactions $H_{Q}$, $E(^{3}D_{1})-E(^{3}D_{2})=5$ cm$^{-1}$. 

Explicit expressions of hyperfine interaction matrix elements are given by \cite{15,16}
\begin{widetext}
\begin{equation}
\begin{aligned}
   \langle l_{1}l_{2}SLJIF|H_{\mu}|l_{1}l_{2}S'L'J'IF\rangle=&(-1)^{J'+I+F}\sqrt{I(I+1)(2I+1)}\left \{ \begin{matrix}
J&I&F \\
I&J'&1 \\
\end{matrix} \right\}\\
&\times\left(l_1l_2SLJ\left\Vert\sum_{i=1}^{2}a_{l_{i}}[\bm{l}^{(1)}-\sqrt{10}(\bm{\texttt{s}}^{(1)}\times \bm{\texttt{C}}^{(2)})^{(1)}]+a_{s_{i}}\delta_{l_{i},0}\bm{\texttt{s}}^{(1)}\right\Vert l_1l_2S'L'J'\right)
\end{aligned}
\end{equation}
where
\begin{equation}
\begin{aligned}
   &\left(l_1l_2SLJ\left\Vert\sum_{i=1}^{2}a_{l_{i}}\bm{l}^{(1)}\right\Vert l_1l_2S'L'J'\right)=(-1)^{S+L'+J+l_{1}+l_{2}}\sqrt{(2J+1)(2J'+1)(2L+1)(2L'+1)}\left \{ \begin{matrix}
L&J&S \\
J'&L'&1 \\
\end{matrix} \right\}\\
 &\qquad\quad\times\left[\langle a_{l_{1}}\rangle(-1)^{L'}\sqrt{l_{1}(l_{1}+1)(2l_{1}+1)}\left \{ \begin{matrix}
l_{1}&L&l_{2} \\
L'&l_{1}&1 \\
\end{matrix} \right\}+\langle a_{l_{2}}\rangle(-1)^{L}\sqrt{l_{2}(l_{2}+1)(2l_{2}+1)}\left \{ \begin{matrix}
l_{2}&L&l_{1} \\
L'&l_{2}&1 \\
\end{matrix} \right\}\right]\delta_{S,S'}
\end{aligned}
\end{equation}

\begin{equation}
\begin{aligned}
 &\left(l_1l_2SLJ\left\Vert\sum_{i=1}^{2}a_{l_{i}}(\bm{\texttt{s}}^{(1)}\times\bm{\texttt{C}}^{(2)})^{(1)}\right\Vert l_1l_2S'L'J'\right)\\
  &\qquad\qquad\qquad\quad=(-1)^{l_{1}+l_{2}}\frac{3}{\sqrt{2}}\sqrt{(2J+1)(2J'+1)(2L+1)(2L'+1)(2S+1)(2S'+1)}\left \{ \begin{matrix}
\frac{1}{2}&S&\frac{1}{2} \\
S'&\frac{1}{2}&1 \\
\end{matrix} \right\}\left \{ \begin{matrix}
S&S'&1 \\
L&L'&2 \\
J&J'&1\\
\end{matrix} \right\}\\
&\qquad\qquad\qquad\qquad\qquad\qquad\times\left[\langle a_{l_{1}}\rangle(-1)^{S'+L'}(l_{1}\Vert\bm{\texttt{C}}^{(2)}\Vert l_{1})\left \{ \begin{matrix}
l_{1}&L&l_{2} \\
L'&l_{1}&2 \\
\end{matrix} \right\}+\langle a_{l_{2}}\rangle(-1)^{S+L}(l_{2}\Vert\bm{\texttt{C}}^{(2)}\Vert l_{2})\left \{ \begin{matrix}
l_{2}&L&l_{1} \\
L'&l_{2}&2 \\
\end{matrix} \right\}\right]
\end{aligned}
\end{equation}

\begin{equation}
\begin{aligned}
   \left(l_1l_2SLJ\left\Vert\sum_{i=1}^{2} a_{s_{i}}\delta_{l_{i},0}\bm{\texttt{s}}^{(1)}\right\Vert l_1l_2S'L'J'\right)=&(-1)^{S+L+J'+1}\sqrt{\frac{3}{2}}\sqrt{(2J+1)(2J'+1)(2S+1)(2S'+1)}\\
   &\times\left \{ \begin{matrix}
S&J&L \\
J'&S'&1 \\
\end{matrix} \right\}\left \{ \begin{matrix}
\frac{1}{2}&S&\frac{1}{2} \\
S'&\frac{1}{2}&1 \\
\end{matrix} \right\}[(-1)^{S'}\langle a_{s_{1}}\rangle\delta_{l_{1},0}+(-1)^{S}\langle a_{s_{2}}\rangle\delta_{l_{2},0}]\delta_{L,L'}
\end{aligned}
\end{equation}
\end{widetext}

and
\begin{widetext}
\begin{equation}
\begin{aligned}
 \langle l_{1}l_{2}SLJIF|&H_{Q}|l_{1}l_{2}S'L'J'IF\rangle\\
& =(-1)^{J'+I+F+1}\sqrt{\frac{(2I+3)(2I+1)(I+1)}{4I(2I-1)}}\left \{ \begin{matrix}
J&I&F \\
I&J'&2 \\
\end{matrix} \right\}
\left(l_1l_2SLJ\left\Vert e^{2}Q\sum_{i=1}^{2}r_{i}^{-3} \bm{\texttt{C}}^{(2)})\right\Vert l_1l_2S'L'J'\right)
\end{aligned}
\end{equation}
where
\begin{equation}
\begin{aligned}
 &\left(l_1l_2SLJ\left\Vert e^{2}Q\sum_{i=1}^{2}r_{i}^{-3} \bm{\texttt{C}}^{(2)})\right\Vert l_1l_2S'L'J'\right)\\
& \qquad\qquad\qquad\qquad\qquad\quad=(-1)^{S+L'+J+l_{1}+l_{2}}\sqrt{(2J+1)(2J'+1)(2L+1)(2L'+1)}\left \{ \begin{matrix}
L&J&S \\
J'&L'&2 \\
\end{matrix} \right\}\\
&\qquad\qquad\qquad\qquad\qquad\quad\times\left[\langle b_{l_1}\rangle(-1)^{L'} (l_1\Vert \bm{\texttt{C}}^{(2)})\Vert l_1)\left \{ \begin{matrix}
l_1&L&l_2 \\
L'&l_1&2 \\
\end{matrix} \right\}+\langle b_{l_2}\rangle(-1)^{L} (l_2\Vert \bm{\texttt{C}}^{(2)})\Vert l_2)\left \{ \begin{matrix}
l_2&L&l_1 \\
L'&l_2&2 \\
\end{matrix} \right\}\right]\delta_{S,S'}
\end{aligned}
\end{equation}
\end{widetext}

To calculate matrix elements, we reduce to the calculation of single-particle matrix elements $\langle a_{l}\rangle$ and $\langle b_{l}\rangle$. For the 5$s$6$d$ configuration, the expression of $\langle a_{6d}\rangle$ and $\langle b_{6d}\rangle$, according to Sobel'man, can be well approximated by introducing $a_{5s}$, given by \cite{28, 15}
\begin{eqnarray}
    \langle a_{6d}\rangle=\frac{3}{8}\frac{a_{5s}}{l(l+1)(l+\frac{1}{2})}\left(\frac{\varepsilon_{6d}}{\varepsilon_{5s}}\right)^{3/2}\label{eq:b8}\\
     \langle b_{6d}\rangle=\frac{3}{4}\frac{Q}{a_{0}^{2}}\frac{a_{5s}}{\alpha^{2}g_{I}\frac{m}{m_{p}}}\frac{1}{l(l+1)(l+\frac{1}{2})}\left(\frac{\varepsilon_{6d}}{\varepsilon_{5s}}\right)^{3/2}\label{eq:b9}
\end{eqnarray}
where $g_{I}$=0.2428, $a_{0}=5.2917\times10^{-11}$ m, $Q=0.335\times10^{-28}$ m$^{2}$, $a_{5s}$= -1001(2) MHz; $\varepsilon_{5s}$ and $\varepsilon_{6d}$ are the binding energies of the 5$s$ and 6$d$ orbits \cite{29} are 5.69 eV and 0.77 eV, respectively. By taking these values into Eqs.~(\ref{eq:b8}) and (\ref{eq:b9}), $\langle a_{6d}\rangle$ and $\langle b_{6d}\rangle$ are calculated to be -1.25 MHz and -1.26 Hz, respectively. The values for \{$l_{1}, l_{2}, S, S', L, L', J, J', I$\} are assigned to be \{0, 2, 1, 1, 2, 2, 1, 2, 9/2\}. The element ($l\Vert \bm{\texttt{C}}^{(2)})\Vert l$)=-$\sqrt{\frac{l(l+1)(2l+1)}{(2l+3)(2l-1)}}$ is calculated for $l_{1}$ and $l_{2}$, i. e., ($l_1\Vert \bm{\texttt{C}}^{(2)})\Vert l_1$)=0, ($l_2\Vert \bm{\texttt{C}}^{(2)})\Vert l_2$)=-1.2. As a result, $\left(l_1l_2SLJ\left\Vert\sum_{i=1}^{2}a_{l_{i}}(\bm{\texttt{s}}^{(1)}\times\bm{\texttt{C}}^{(2)})^{(1)}\right\Vert l_1l_2S'L'J'\right)$=0, $\left(l_1l_2SLJ\left\Vert\sum_{i=1}^{2}a_{s_{i}}\delta_{l_{i},0}\bm{\texttt{s}}^{(1)}\right\Vert l_1l_2S'L'J'\right)$=-1061 MHz and $\left(l_1l_2SLJ\left\Vert\sum_{i=1}^{2}a_{l_{i}}\bm{l}^{(1)}\right\Vert l_1l_2S'L'J'\right)$=-2.65 MHz. For the electric quadrupole interaction, the matrix element $\left(l_1l_2SLJ\left\Vert e^{2}Q\sum_{i=1}^{2}r_{i}^{-3} \bm{\texttt{C}}^{(2)})\right\Vert l_1l_2S'L'J'\right)$ is calculated to be negligible, indicating the magnetic dipole-electric quadrupole interaction to be negligible with respect to the magnetic dipole interaction. Accordingly, the second-order hyperfine splittings are derived to be $\Delta E^{(2)}_{F}=14.4$, 24.1, 21.9 MHz for $F$=11/2, 9/2, 7/2, respectively. The center of gravity of $^{3}$D$_{1}$ is corrected by 20 MHz following the relation of $\Delta E_{\texttt{cog}}=\frac{2}{5}\Delta E_{11/2}+\frac{1}{3}\Delta E_{9/2}+\frac{4}{15}\Delta E_{7/2}$.

\subsection{\label{app:sec82}Experimental evaluation}
The 2D King plot for $\gamma:$ $^{3}$P$_{0}\to^{3}$D$_{1}$ and $\alpha:$ $^{1}$S$_{0}\to^{3}$P$_{1}$ transitions before 
\begin{figure}
\centering
\includegraphics[width=8 cm]{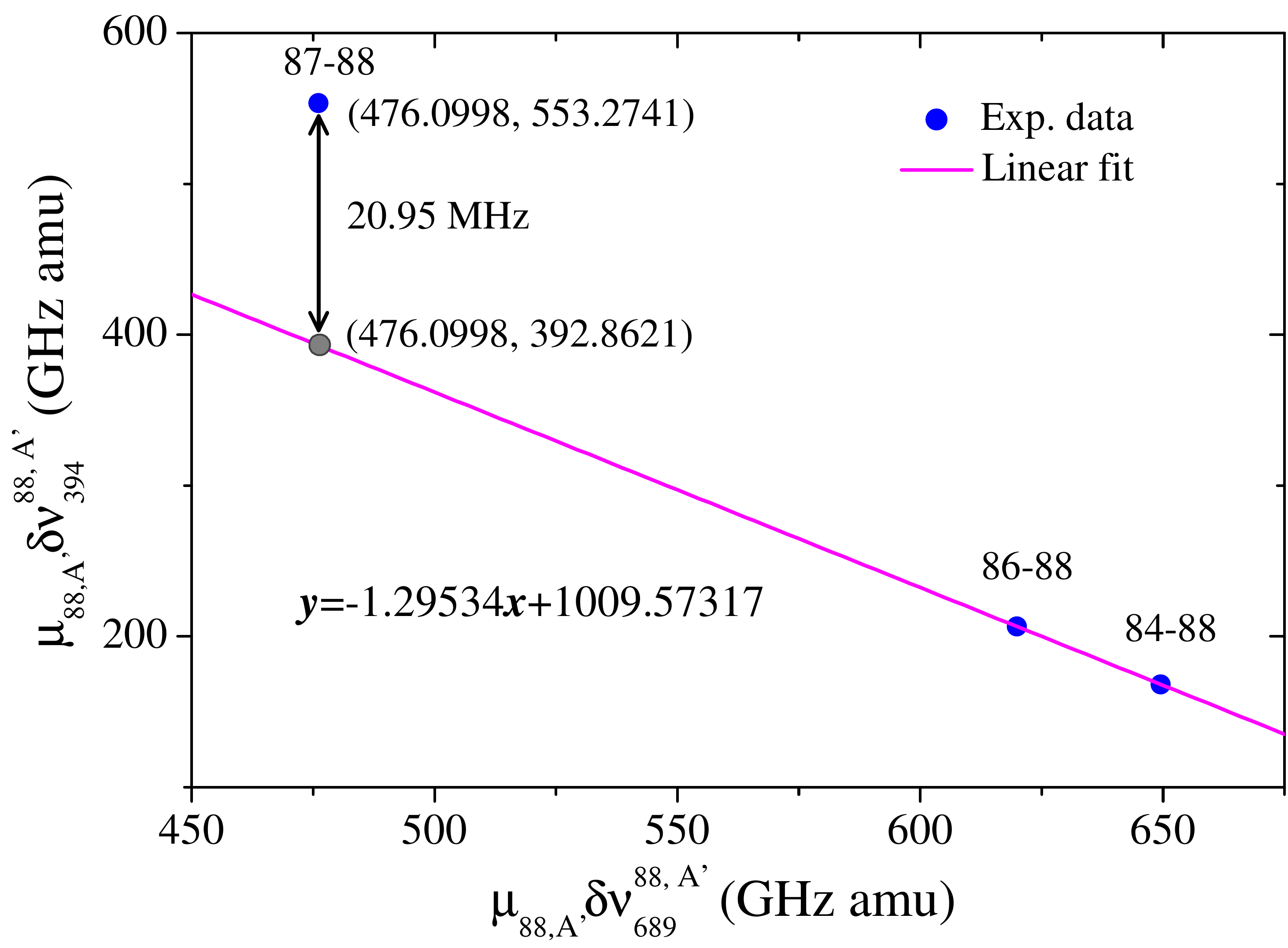}
\captionsetup{justification=raggedright}
\caption{\label{fig:0002} 2D King plot of $^{3}$P$_{0}\to^{3}$D$_{1}$ and $^{1}$S$_{0}\to^{3}$P$_{1}$ transitions before correcting the second-order contribution to the isotope shift of $^{87}$Sr. The blue dots represent the experimental data and the magenta line is linear fit to the even isotope shifts only. The gray dot for $^{87}$Sr is predicted from the fitted line. The difference of modified IS between the experimental measurement and the prediction is 160.412 GHz amu, corresponding to 20.95 MHz frequency shift, which is primarily due to the second-order contribution. }
\end{figure}
correcting the second-order contribution to the isotope shift of $^{87}$Sr is shown in Fig.~\ref{fig:0002}. The King plot line is determined by a linear fitting to even isotope shifts only. The odd modified IS data point is plotted which sits far away from the straight line as a result of the second-order hyperfine interaction on the $^{3}$D$_{1}$ state. 
The reference transition $\alpha$ is free from second-order hyperfine interactions within limits of error.To extract second-order hyperfine correction of the isotope shift on the $\gamma$ transition for $^{87}$Sr, we calculate the difference between the experimentally measured IS and the predicted value from King plot line determined only by even isotope shifts. Based on the fitted line, the modified IS of $^{87}$Sr of the $\gamma$ transition is predicted to be 392.8621 GHz amu, which is much less than the experimental measurement result of 553.2741 GHz amu. The difference of modified IS 160.412 GHz amu reveals the second-order contribution of 20.95 MHz, which is consistent with the theoretically calculated result of 20 MHz. The second-order contribution here refers to the shift of the centre of gravity of $^{3}$D$_{1}$ due to second-order hyperfine interaction.




\nocite{*}
\bibliography{apssamp}

\end{document}